# A Soft Vibrotactile Display Using Sound Speakers

Keitaro Ihara[1], Atsuya Baba[1], Hiroki Ishizuka[2], Takefumi Hiraki[3], and Osamu Oshiro[2]

[1] *School of Engineering and Science. Osaka University, Osaka, Japan*

[2] *Graduate School of Engineering and Science. Osaka University, Osaka, Japan*

[3] *Cluster Metaverse Lab, Tokyo, Japan*

(Email: ishizuka@bpe.es.osaka-u.ac.jp)

**Abstract ---** This study introduces an innovative vibrotactile display that harnesses audio speakers to convey tactile information to the fingertips while preserving the display's softness and flexibility. Our proposed system integrates a flexible polymer body with silicone rubber tubes connected to audio speakers. By streaming audio through these speakers, we induce air vibrations within the tubes, generating tactile stimuli on the skin. In contrast to conventional tactile displays that often rely on bulky, rigid actuators, our approach employs multiple speakers to deliver high-resolution vibration patterns. This configuration enables the presentation of high-frequency vibrations, potentially enhancing the fidelity of tactile feedback. We present a detailed description of the display's design principles and implementation methodology, highlighting its potential to advance the field of haptic interfaces.

## 1 INTRODUCTION

Tactile sense is a crucial sense that can confirm the existence of objects. The widespread adoption of recording and reproducing tactile information could lead to the development of technologies that virtually recreate the presence of objects. This technology will enhance remote operations for robots, enable the transmission of tactile sensation of objects in online shopping, and improve the realism in virtual environments. Tactile displays are being researched as devices to reproduce tactile information. Among these, vibration tactile displays can be used to reproduce the texture of an object's surface by reproducing the vibration patterns. The principles behind vibration tactile displays include linear resonant actuators [1], piezoelectric actuators [2], and shape memory alloy actuators [3]. However, these typically require large actuators, limiting spatial resolution of tactile feedback. Additionally, actuator rigidity can impede natural user movements.

In this study, we propose a tactile display using an audio speaker. This tactile display delivers tactile information to the fingertips by transmitting sound waves output by the speaker through a soft device with an internal channel. While several tactile displays using audio speakers have been proposed [4, 5], they typically use a single speaker to present tactile information. In contrast, our research utilizes multiple audio speakers to present high-resolution vibration information. This paper explains the principles behind this approach and describes the device's configuration.

## 2 PRINCIPLE

Figure 1 illustrates the principle of the proposed vibrotactile display. The device consists of a soft polymer body with holes to which silicone rubber tubes connected to a speaker are attached. Streaming audio through the speakers vibrates the air inside the tube while a fingertip covers the holes, causing skin vibration and presenting a vibrotactile stimulus to the user. This principle is similar to pneumatic vibrotactile displays. However, traditional pneumatic displays are limited by valve switching times, making it difficult to present high-frequency stimuli [6].

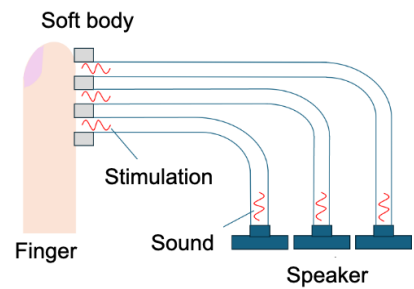

Fig. 1 Proposed method.

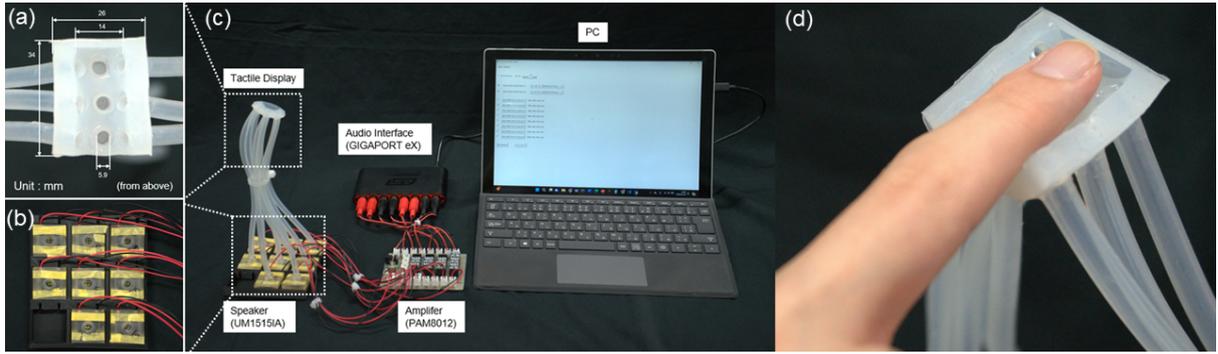

Fig. 2 Implementation of the proposed vibrotactile display. (a) Body and its dimension. (b) Speaker module. (c) Setup for tactile stimulation. (d) Actual use case.

Our proposed display uses an audio speaker capable of generating high-frequency pneumatic pressure waves, enabling a broader frequency spectrum. While the presentable vibration frequencies depend on speaker characteristics, our system demonstrates favorable performance across a wide bandwidth from a few Hz to several hundred Hz. We have empirically verified the capability to present vibrations up to several hundred Hz. Additionally, since this method directly uses the speaker's sound, recorded audio can be employed straightforwardly as a tactile stimulus, simplifying signal generation.

## 3 IMPLIMENTATION

This section explains the implementation method of the proposed vibrotactile display. The body was fabricated using a soft lithography process. First, a plastic mold was printed using a photopolymer 3D printer (Form 3, Formlabs). Then, silicone rubber (Ecoflex 0030, Smooth-On) was poured into the mold and cured thermally to form the main body of the display. The display incorporates silicone rubber tubes (5.9 mm inner diameter) inserted into holes (Fig. 2(a)). An audio speaker (UM1515IA, DB Products) is connected to this tube (Fig. 2(b)). The audio signal to the speaker is amplified by an amplifier (PAM8012, Diodes) from a signal provided by an audio interface (GIGAPORT eX, ESI) connected to a computer (Fig. 2(c)). Each speaker outputs sound according to the computer's audio data, and the vibrotactile display's corresponding location provides the vibrotactile stimulus. Users can experience vibrational tactile stimulation by placing their fingers on the haptic display (Fig. 2(d)). At this time, it is necessary for the tube's interior to be sealed by the skin.

## 4 Conclusion

This paper aimed to realize a soft vibrotactile display capable of presenting high-frequency vibration of pneumatic pressure. A soft vibrotactile display capable of presenting high-frequency vibrations was achieved by utilizing the air vibrations produced by the audio speaker. In the future, the structure of this vibrotactile display will be examined to explore better designs. Additionally, an investigation will be conducted on how the vibrotactile stimuli presented by this display are perceived.

References

[1] A. Dementyev, A. Olwal and R. F. Lyon: "Haptics with Input: Back-EMF in Linear Resonant Actuators to Enable Touch, Pressure and Environmental Awareness," Proceedings of UIST '20, pp. 420-429 (2020).

[2] R. Schweizer, D. Voit and J. Frahm: "Finger representations in human primary somatosensory cortex as revealed by high-resolution functional MRI of tactile stimulation," NeuroImage, Vol. 42, No. 1, pp. 28-35 (2008).

[3] J. Xu, Y. Kimura, K. Tsuji, K. Abe, T. Shimizu, H. Hasegawa and T. Mineta: "Fabrication and Characterization of SMA Film Actuator Array with Bias Spring for High-Power MEMS Tactile Display," Microelectronic Engineering, Vol. 227, 111307, 2020.

[4] Y. Hashimoto, S. Nakata, and H. Kajimoto: "Novel tactile display for emotional tactile experience," Proceedings of ACE '09, pp. 124-131 (2009).

[5] M. Ito, R. Sakuma, H. Ishizuka and T. Hiraki: "AirHaptics: Vibrotactile Presentation Method using an Airflow from Audio Speakers of Smart Devices," Proceedings of VRST '22, 2 pages (2022).

[6] B. Shan, C. Liu, Y. Guo, Y. Wang, W. Guo, Y. Zhang and D. Wang: "A Multi-Layer Stacked Microfluidic Tactile Display With High Spatial Resolution," IEEE Transactions on Haptics (Early Access).